\documentstyle[12pt]{article}
\title{Effective   quark
Lagrangian in the instanton--gas model}
\author{Yu.A.Simonov\\
Institute of Theoretical and Experimental Physics\\ 117259,Moscow
, B.Cheremushkinskaya 25, Russia}
\newcommand{\be}{\begin{equation}}
\newcommand{\ee}{\end{equation}}
\begin{document}
\maketitle

\begin{abstract}

A straightforward derivation of the effective quark
Lagrangian is presented for the topologically neutral chiral broken
phase of the dilute instanton gas. The resulting quark Lagrangian is
a nonlocal NJL -- type and contains 4q, 6q,... vertices for any
number of flavours. Correspondence of this result with
previously  known in literature is discussed in detail.

\end{abstract}

\section{Introduction}

The instanton gas model has a long  history [1,2,3]. Recently lattice
measurements of quark and gluon correlators  in coordinate space [4]
have revealed important contributions of
 instantons, which gives an additional impetus to the
model [5].

On the theoretical side the instantons have been proved to solve the
$U(1)$ problem [6], and to generate a specific effective quark
Lagrangian  (EQL) on a net  topological charge [7].

In the topologically neutral system, such as the QCD vacuum,
instantons have  been proposed as a driving mechanism for the chiral
symmetry breaking (CSB) [1,2,8]. The quantitative theory of  light
quarks (TLQ) in the instanton gas has been suggested in [9] and later
elaborated in  [10,11,12].

There are a considerable number of papers which use the
method for quantitative predictions and comparison with experiment
(as an example see [13]).

The essential ingredient of TLQ is the assumption of the  dominant
role of the   quark zero modes on each instanton, and the resulting
EQL is proposed
in the form of  the
$2N_f\times 2N_f$ determinant of quark fields with coefficients
(vertices) $M(p,p')$ expressed through quark zero modes.

As a result TLQ predicts CSB and the creation of chiral quark mass at
all instanton densities; after bosonization the  effective bosonic
Lagrangian was derived [9, 10] containing expected Goldstone bosons.

In view of numerous  successful applications we believe it is
important to test the accuracy of the  method.

In its modern form [12] the TLQ contains some approximations, first of
all the zero-mode approximations (ZMA) introduced already in [9],
and the basic role in TLQ is played  by  the ansatz for EQL
[10-12] of the 'tHooft form [7].
The ZMA was questioned in [14], and it was argued there that nonzero
modes play an important role in addition to zero modes.

In the present paper a new method
 is proposed, where all modes are accounted for
automatically and we are avoiding standard approximations of TLQ.

In the  next  section we demonstrate a
straightforward and exact way to derive the EQL
 in the topologically neutral dilute instanton gas.
 Our results differ from the existing in several points  and
 lay ground for a new formalism of light quarks in the instanton
gas.

 First of all, our EQL contains a set of quark terms starting
from 4q for any number of flavours. The 4q term differs  from that
of TLQ and is similar instead to the extended NJL type Lagrangian
[15].  The structure of our EQL makes it possible to create chiral
mass and CSB, and bosonization yields the Nambu-Goldstone boson
Lagrangian.

All coefficients in our EQL are explicitly known and
this enables one
to do  quantitative predictions. In particular the appearance of
the 4q vertex for any $N_f$ is important from the point of view of
theory (such as effective top-condensate model of electroweak
interactions [16]) and applications (OZI violation vertex, quark--pair
creation, strangeness mixture).

In  our formalism  there are several ways to find
the chiral quark mass entering the averaged quark propagator. One way
exploits Dyson--Schwinger equations for our EQL in connection with
the mean--field (Hartree) approximation.
This is equivalent to the direct summing of leading trms in the
$1/N_c$ expansion. In the one--flavour case one obtains an integral
equation for the chiral mass, which coincides with the equation
obtained earlier in [17] in a different, diagrammatic way.  However,
instead of the zero-mode approximation made in [17], one can  reduce
this equation to the one--dimensional integral one, subject to a
simple numerical analysis.

This is done in section 3 of the paper. In case of two and more
flavours one should use instead the bosonization method and find the
chiral mass and chiral condensate from the steepest descent method,
which will be a subject of the next paper.

Finally in the last section a detailed discussion is given of the
correspondence of our method with TLQ.

 \section{Effective quark
Lagrangians}

We consider in this letter the neutral dilute gas of  $N_+$
instantons $I$ and $N_-$ antiinstantons $\bar I$, $N_+=N_-=N/2$, with
the superposition principle [1-3]
\be
A_{\mu}(x)=\sum^N_{i=1}A_{\mu}^{(i)}(x,\gamma)
\ee
where $\gamma$ denote the set of collective coordinates for each
instanton $\gamma_i=\{R^{(i)}, \Omega^{(i)}, \rho^{(i)}\}$. In what
follows we for simplicity fix the instanton size  $\rho^{(i)}=\rho$
and neglect perturbative gluon effects  together with $II, I\bar I$
interaction, assuming that the latter stabilizes gas at small
density.

Then the effective quark Lagrangian $L_{EQL}$ can be defined as
$$
Z=\int d\gamma D\psi D\bar{\psi}e^{\sum^{N_f}_f\int
\bar{\psi}_fS^{-1}\psi_f dx}=
$$
\be
=
\int  D\psi D\bar{\psi}e^{-\sum^{N_f}_f\int
(\bar{\psi}_fS^{-1}_0\psi_f) dx+ {\cal{L}}_{EQL}(\psi,\bar{\psi})}
\ee
where
\be
S^{-1}\equiv -i\hat{\partial}-im_f-g\sum_{i=1}^N\hat
A^{(i)}(x,\gamma)
\ee
\be
S^{-1}_0= -i\hat{\partial}-im_f
\ee
and
\be
d\gamma\equiv \prod^N_{i=1}(d\Omega_i\frac{d^4R^{(i)}}{ V_4})
\ee
The resulting $L_{EQL}$ is a polynomial of $\psi, \bar{\psi}$
of infinitely large power.

To obtain $L_{EQL}$ one has to average in (2) over $d\gamma$ (5),
$\gamma$ entering into $A^{(i)}_{\mu}$ as
\be
A^{(i)}_{\mu}(x,\gamma)=\bar{\eta}_{a\mu\nu}
\frac{(x-R^{(i)})_{\nu}\rho^2\Omega^+_i\tau_a\Omega_i}
{(x-R^{(i)})^2[(x-R^{(i)})^2+\rho^2]}
\ee

The straightforward use of the cluster expansion [18] for the
averaging over $d\gamma$ yields
\be
Z=\int D\psi D\bar{\psi}e^{-\sum^{N_f}_f
(\bar{\psi}_fS^{-1}_0\psi_f) dx+
\sum^N_{i=1}\sum^{\infty}_{n=1}\frac {1}{n!}\ll\theta^n_i\gg}
\ee
where we have denoted ($f_i$ -flavour indices).
\be
\theta_i=\sum_{f_k=1}^{N_f}\theta^{f_k}_i,
\theta^{f_1}_i \equiv \int dx \psi^+_{f_1}(x)\hat
A^{(i)}\psi_{f_1}(x);
\ee
 and angular brackets imply integration over
$d\gamma$, while double brackets stand for cumulants [18] e.g.
  \be
\ll\theta^2\gg_{\Omega_iR_i}=<\theta^2>_{\Omega_iR_i}-<\theta>_{\Omega_iR_i}
<\theta>_{\Omega_iR_i}
\ee

Let us look at $\ll \theta^n_i\gg$ more closely. The first term with
$n=1$ vanishes due to the integration  over $d\Omega_i$.
The next term with $n=2$ will be of most interest to us
in what follows. This can be written in the form (see
Appendix 1 for details of derivation)
$<\theta^2_i>=\sum_{f_1f_2}<\theta^{f_1}_i\theta ^{f_2}_i>$, and
finally cast in the form,
$$
\sum_i<\theta^2_i>=
\int
\frac{dpdp_1dq}{(2\pi)^{12}}
\Gamma(q)\{(1-\frac{1}{N^2_c})\sum_{i,F}c_iJ_i^{F}(p)
J_i^{F}(p_1)+
$$
$$
+\frac{1}{2N_f}(1-\frac{1}{N^2_c})\sum_{i}c_iJ_i^{0}(p)
J_i^{0}(p_1)-\frac{1}{q^2}\sum_{i,F}c'_i\tilde J^{F}_i(p)\tilde
J_i^{F}(p_1)-
 $$
\be
-\frac{1}{2q^2N_f}\sum_{i}c'_i\tilde J_i^{0}(p)
\tilde
J_i^{0}(p_1)\}
\ee
where $c_i,c'_i$ are given in Appendix 1, and
$$
J_i^{n}(p)=\psi^+(p)t^n0_i\psi(p_1-q),
~J^n_i(p_1)=\psi^+(p_1)t^n0_i\psi(p+q),
 $$
\be
\tilde
J_i^{n}(p)=\psi^+(p)\hat qt^n0_i\psi(p_1-q),
~\tilde J^n_i(p_1)=\psi^+(p_1)\hat q t^n0_i\psi(p+q)
\ee
and
 $n=F$ or $0$, with $F=1,...
N_f^2-1,$  and $t^0=\hat 1, t^F$ is the flavour group generator.

We have also defined in (10)
\be
\Gamma(q)=\frac{4q^2\varphi^2(q)(4\pi^2\rho^2)^2N}{(N^2_c-1)V_4},
\ee
and $\varphi(q)$ is the Fourier transform of the instanton field (6)
given in Appendix 1, eq (A.6) .

The 4q interaction term (10) is of the extended NJL type, containing
all possible nonlocal bilinears.

The advantage of  (10) over the NJL  Lagrangian is that it provides
the in-built formfactor $\Gamma(q)$ which makes loop corrections
finite since it behaves as
\be
\Gamma(q)\approx \frac{N}{V_4}\frac{(4\pi^2\rho^2)^2}{(N^2_c-1)}
\left \{
\begin{array}{ll}
\frac{1}{q^2},&q\to 0\\
\frac{16}{\rho^4q^6},& q\to \infty
\end{array}
\right .
\ee
One can see in (13) that the small $q$ behaviour of $\Gamma(q)$
(which is due to
the $1/x^3$      asymptotics of the instanton vector potential(6))
implies massless exchanges between currents $J(q)$ at large distances
-- a certainly unphysical feature, which is due to the ansatz (1) of
noninteracting instanton gas. Taking into account the $II$ and
$I\bar I$ interactions changes the instanton profile at large
distances (possible forms are suggested in [3])
and makes $\Gamma(q\to 0)$ finite in the realistic instanton gas or
liquid model.

There are several important properties of the new  EQL (7).

First of
all, the sum over $n$ is not limited, so therefore it contains
vertices containing 4q,6q,... etc to infinity for any $N_f$. This
differs from the standard ansatz of [9-12], where only the term with
$2N_f$ quark operators appears.

One can  observe in (8) the contribution of all quark modes,
since one can rewrite $\theta^i$ as follows
\be
\theta_i=\int dx \psi^+(x)\hat A^i(x)\psi(x)=-
\sum_{n,k}<\psi^+|S_0^{-1}|u^i_n>\varepsilon^i_{nk}<u^i_k|S_0^{-1}|\psi>
\ee
with e.g. $<\psi^+|S^{-1}_0|u^i_n>\equiv
\int\psi^+(x)(-i\hat{\partial}-im)u^i_n(x)d^4x$.
 and
 $\varepsilon^i_{nk}$ defined as
 \be
 \varepsilon^i_{nk}=-<u^i_n|S_0 \hat A^{(i)}S_0|u^i_k>
 \ee

 The standard TLQ prescription (9-12) obtains when one  keeps in
 (14) only the zero mode coefficient, $\varepsilon^i_{00}$,

Consider  now all  terms in (7) with $n\geq 2$ and take the limit
$N_c\to \infty$ to perform the  averaging over $d\Omega$ explicitly.

The EQL (7) is additive for instantons, therefore we consider it for
the $i$-th instanton and omit the subscript $i$ in what follows.  One
has \be
{\cal{L}}_i=\sum^{\infty}_{n=2}\frac{1}{n!}\ll\theta^n_i\gg,~~~
 {\cal{L}}=\sum_i{\cal{L}}_i
  \ee

  The double brackets in (16) imply  subracting from $<\theta^n>$
  disconnected pieces which appear while averaging in $d\Omega dR$.

  Since each integration $dR$  is accompanied by $1/V_4$ the
  second term in the difference,
  \be
  <\theta^n>_R-<\theta^k>_R<\theta^{n-k}>_R,
  \ee
  vanishes in the   thermodynamical limit, $N\to \infty,~~ V_4\to
  \infty, ~~ \frac{N}{V_4} =const$. Hence the only terms left in $\ll
  \theta^n\gg$ are of the form
  \be
  <\theta^n>_{R,\Omega}=<<\theta^n>_{\Omega}-<\theta^2>_{\Omega}
  <\theta^{n-2}>_{\Omega}-....>_R
  \ee
  Now in averaging over  $d\Omega$ one obtains in the leading terms
  of the $1/N_c$ expansion the set of disconnected (in color indices)
  products, which are subtracted in (18) and $(n-1)!$  equivalent
  connected terms, which differ only by permutation of indices. Thus
  one comes to the EQL defined in (16)

$$
{\cal{L}}=\sum^{\infty}_{n=2}\frac{N}{2n V_4N_c^n} \prod^n_{k=1}\int
\psi^+_{\alpha_k}(p_k)\gamma_{\mu_k}\psi_{\beta_k}(p_k-q_k)
\frac{dp_kdq_k}
{(2\pi)^8}\times
$$
\be
\times (2\pi)^4\delta^4(\sum^n_{i=1}q_i) tr
(\prod^n_{k=1}A^I_{\mu_k}(q_k))
\prod^n_{i=1}\delta_{\alpha_i, \beta_{i-1}}+(I\to \bar I)
\ee

The EQL (7) and (19) is the central point of this letter and in the
following section we use it to derive equations for the chiral mass.

 \section{ Chiral mass, chiral symmetry breaking and the gap equation}

The multiquark terms $\ll \theta^n\gg$ in the effective Lagrangian
(7) contain all information about possible quark mass creation which
automatically signifies CSB. The form of $<\theta^2 >$ in (10)  tells
us that the averaged 4q interaction is of the extended NJL type [15]
and is nonlocal.

Nevertheless one can use for the 4q term  the same gap equation,
derived from the Dyson--Schwinger method, as it was done in the NJL
formalism [15]. However the essential difference of our expression (7)
from the NJL Lagrangian is that the former contains all powers $n$,
i.e. all n--quark effective vertices, $n=4,6,...\infty$ .

Therefore the gap equation, defining the creation of the chiral mass
in our case, should contain in general this infinite number of
vertices, unless one can find a small parameter suppressing terms
with large  $n$. To answer this question we shall consider first the
case of one flavour and compute the averaged quark propagator in the
instanton gas.

We shall  derive it in the  large $N_c$ limit
 and shall
obtain the integral equation for the chiral mass, containing
contributions of all n--quark vertices.

At this point one can find from (19) the effective quark mass in the
mean--field (Hartree) approximation. To this end one should identify
in $L$ (19) the structure $-i\int
\frac{dp}{(2\pi)^4}\psi^+(p)M(p)\psi(p)$. In the leading order at
large $N_c$ one replaces in (19) $(n-1)$ colorless  bifermion
combinations by their vacuum expectation values, and in Hartree
approximation by the effective quark propagator $\bar S(p)$
\be
\bar S(p)=\frac{\hat 1_c}{\hat p-iM(p)}
\ee
Thus each of the colorless bifermions yields a factor $N_c$ and we
obtain
\begin{eqnarray}
\nonumber
M(p)&=& -i\frac{N}{2V_4N_c}\sum^{\infty}_{n=2}\int
\prod^n_{k=1}(\frac{dq_k}{(2\pi)^4})(2\pi)^4\delta^4(\sum^n_{i=1}q_i)
tr(\hat A (q_1)
\bar S(p-q_1)...\hat A (q_n))+\\
&+&
(I\to \bar I)
\end{eqnarray}
where the last bracket in (21) refers to the contribution of
antiinstantons.

One can rewrite this expression in a simpler form

\be
M(p)=+i\frac{1}{N_cV_4}\sum_i<p|(\bar S-(\hat A^i)^{-1})^{-1}|p>_R
\ee

Eq.(22) is equivalent to Eq.(7) of [17] when one does Fourier
transformation and averaging over $dR_i$.

Since $\bar S(p)$ contains $M(p)$ as in (20), two coupled equations
(20) and (21) define $M(p)$ through $\hat A(p) $ and the density
parameter $\frac{N}{N_cV_4}$.

One can introduce the "Green's functions" $G(\bar G)$ for the quark
propagation in the in the instanton (antiinstanton) field $A(\bar A)$
with effective mass $M(p)$, e.g.
\be
(\hat p- iM-\hat A) G(p,p')= (2\pi)^4\delta(p-p')
\ee

Separating $\delta$--function term from $G$, one obtains
\be
G(p,p')=\frac{(2\pi)^4\delta(p-p')}{\hat p- iM(p)}+
\frac{\xi(p,p')}{(\hat p-iM(p))(\hat p'-iM(p'))}
\ee
and similarly for $\bar G$ and $\bar {\xi}$. Insertion of (24) into
(22) yields
$$
M(p)=-\frac{iN}{2N_cV_4}tr\int\frac{d^4q}{(2\pi)^4}\{\hat A (p-q)
\frac{1}{\hat q -iM(q)}\xi(q,p)+
$$
  \be
  +(A\to\bar A,\xi\to\bar{\xi})\}
  \ee
  and for $\xi$ one has an equation following from (24)
  \be
  \xi(p,p')=\hat A(p-p')+\int\hat A(p-q)\frac{1}{\hat q-
  iM(q)}\xi(q,p')\frac{d^4q}{(2\pi)^4},
  \ee
  and  similar equation for $\bar {\xi}(p,p')$ with $A_{\mu}$
  replaced by the antiinstanton field $\bar A_{\mu}$.

Since one expects $M(0)$ to be around  200-300 MeV, [9-12, 17], the
spectral expansion of $G$ and the ZMA done in [17] which is valid for
vanishing $M(0)$ is not convincing. Therefore one has to solve  Eqs.
(25-26) numerically.

 The calculations done above in this section, demonstrate that (19)
 is the EQL corresponding to the Pobylitsa   equation [17] obtained
 for $N_f=1$ and $N_c\to \infty$. In case of $N_f>1$ the EQL is given
 in (7) and its explicit form for $n=2$ in (10). For many flavours
 $(N_f\geq 2)$ it is natural and (easier) to use the bosonized form
 of EQL, which will be published elsewhere.

Finally we discuss in the section symmetries of our EQL,
$\sum_n\frac{\ll\theta^n\gg}{n!}$. We consider only the case
$N_+=N_-$, so that the net topological charge of the system is zero,
and there are no global zero modes. Correspondingly the effective
'tHooft Lagrangian [7] with the global zero modes does not occur in
this case. Now the 4q term in (7) can be
rewritten as in (A.9), the form which respects  $U(N_f)\times U(N_f)$
symmetry.  The same can be shown for all terms in (7), and therefore
violation of $SU(N_f)\times SU(N_f)$ can occur only
spontaneously while $U_A(1)$ has anomaly. The explicit example of
CSB is given by a nonzero solution of (25-26),  derived above
in the limit $N_c\to \infty$.

 \section{ Summary and discussion
 }

 The main result of the letter is the EQL (7), with explicit form for
 $n=2$ and any $N_f$ in (10), and for all $n$  and $N_f=1$, $N_c\to
 \infty$ in (19).

 The 'tHooft--type ansatz of EQL of ref. [10-12] is obtained from our
 EQL (7) when one selects there $n=N_f$ and keeps in the mode
 expansion (14) only the zero mode contribution. One can check, that
 in this case automatically appears the determinantal structure of
 the 'tHooft ansatz. We have not found any arguments why nonzero
 modes in (14) and  all terms with $n\neq N_f$ should be dropped.

 It is interesting, that the structure of the 4q term in the EQL Eq.
 (10), is so much resembling the  extended NJL model with all Lorenz
 group channels (see (A.12)), in contrast to the standard TLQ
 Lagrangian [10-12], where in the leading $1/N_c$ order only scalar
 and pseudoscalar  channels appear.

 It is not impossible, however,  that the resulting  effect of all
 terms of EQL in the sum over $n$ in (7),  will be numerically close
 to the standard TQL results [10-12] for  $M(p)$ and $<\bar q q>$. As
 one example of  this sort one may quote the equation (22) for the
 effective quark mass which is obtained above from our EQL
 in the Hartree approximation and which
 coincides with that of ref. [17], where diagrammatic analysis for
 the quark propagator was exploited.

The author of [17] applied the ZMA for (22), valid in the limit of
small $M(p)$, and obtained a solution for $M(p)$ closely resembling
that of the standard TQL analysis [9-12].
However the resulting value of $M(0)\approx 350 MeV$ is not small
and this fact casts some doubts on the whole numerical procedure
of [17].
The crucial check of this coincidence would be the exact
solution of Eqs.(25-26) for $M(p)$ and its comparison with the
standard TQL solution of $M(p)$ [9-12]. This will be  a subject of a
future publication.

As it is, our EQL (7) is an  exact result for  any $N_c$ and any
$N_f$, which enables one to calculate different physical effects,
from $q\bar q$ creation due to the  4q vertex, to chiral symmetry
breaking effects $(M(p),<\bar q q>$ etc).   The corresponding
detailed calculations and comparison with the existing TQL results is
planned for the future and  will help to clarify the  mechanism of
chiral dynamics in the QCD vacuum.

The author is grateful to G.'tHooft for a valuable discussion, to
D.I.Diakonov for a useful discussion and correspondence, and to
B.O.Kerbikov and D.S.Kuzmenko for numerous discussions in the course
of the present work.

This work was supported in part by grants RFFI 96-02-19184, RFFI-DFG
96-02-000 88G and INTAS 94-2851.

 \newpage

\underline{\bf APPENDIX 1 }\\

{\bf Derivation of the 4q term of the  effective  quark
Lagrangian}\\

\setcounter{equation}{0}
\def\theequation{A.\arabic{equation}}

$$
<\theta^2>\equiv  <\int \psi^{+f_1}(x) \hat A(x)\psi^{f_1}(x)\int
 \psi^{+f_2}(y) \hat A(y)\psi^{f_2}(y)dxdy>_{\Omega, R}=
 $$
 \be
 \int(\psi^{+f_1}_{\alpha}(x)\gamma_{\mu_1}\psi^{f_1}_{\beta}(x))dx
 \int(\psi^{+f_2}_{\beta}(y)\gamma_{\mu_2}\psi^{f_2}_{\alpha}(y)dy
 \frac{\Lambda_{\mu_1\mu_2}(x,y)}{N^2_c-1}
  \ee
 where $\alpha,\beta$ are
 color indices, and
 $$
 \Lambda_{\mu_1\mu_2}(x,y)=\frac{2}{V_4}\int d^4R(\delta_{\mu_1\mu_2}
\delta_{\nu\rho}-\delta_{\mu_1\rho}\delta_{\mu_2\nu}
\pm \varepsilon_{\mu_1\nu\mu_2\rho})(x-R)_{\nu}(y-R)_{\rho}
\times
$$
\be
\times a(x-R)a(y-R)
\ee
where
\be
a(x)=\frac{\rho^2}{x^2(x^2+\rho^2)}
\ee
and the sign $+(-)$  refer to the instanton (antiinstanton) case
respectively.

 In (A.1) one can notice that color indices are
intercharged in both quark bilinears, which calls for the application
of Fierz transformation, which gives
 \be
  \gamma_{\mu}\times
\gamma_{\mu}= 1\times 1-\frac{1}{2} (\gamma_{\mu}\times
\gamma_{\mu})-\frac{1}{2} (\gamma_5\gamma_{\mu}\times
\gamma_5\gamma_{\mu}) -\gamma_5\times \gamma_5 \ee Note that we are
using Euclidean $\gamma$ matrices. i.e.  $$ \gamma_4=(
\begin{array}{ll}
1&\\
&-1
\end{array}),
\gamma_i=-i\beta\alpha_i,\gamma_5=(
\begin{array}{ll}
&-1\\
-1&
\end{array}),
\gamma_{\mu}\gamma_{\nu}+\gamma_{\nu}\gamma_{\mu}=2\delta_{\mu\nu}
$$

  For the following it is useful to go over to the momentum space and
  to this end we need the Fourier transform of the instanton vector
  potential
  \be
  A_{\mu a}(x-R)={2}\bar {\eta}_{\mu a\nu}\rho^2 i4\pi^2\int
  \frac{d^4q}{(2\pi)^4}\cdot q_{\nu}\varphi(q)
  e^{iq(x-R)}
  \ee
  where
  \be
  \varphi(q) = \frac {1}{q^2}
  \{
  K_2(q\rho)-\frac{2}{(q\rho)^2}\}
  \ee
  Insertion of (A.5) into (A.1) yields
  $$
  <\theta^2>=\int \frac{dp}{(2\pi)^4}\frac{dp_1}{(2\pi)^4}
  \frac{dq}{2\pi)^4}\frac{2(4\pi^2\rho^2)^2}{(N_c^2-1)V_4}
 (\psi^{+f_1}_{\alpha}(p)\gamma_{\mu_1}\psi^{f_1}_{\beta}(p+q))
 (\psi^{+f_2}_{\beta}(p_1)\gamma_{\mu_2}\psi^{f_2}_{\alpha}(p_1-q)
 $$
 \be
 (\delta_{\mu_1\mu_2}q^2-q_{\mu_1}q_{\mu_2})\varphi^2(q)
 \ee

 Using (A.4) and a similar Fierz relation for scalars
 \be
1\times 1=\frac{1}{4}\{
1\times 1 +\gamma_{\mu}\times \gamma_{\mu}+\frac{1}{2}
\sigma_{\mu\nu}\times \sigma_{\mu\nu}-
\gamma_5\gamma_{\mu}\times \gamma_5\gamma_{\mu}+
\gamma_5\times \gamma_5\}
\ee
one obtains for (A.7)
  $$
  <\theta^2_i>=\int \frac{dpdp_1dq}{(2\pi)^{12}}
  \frac{2(4\pi^2\rho^2)^2\varphi^2(q)}{(N_c^2-1)V_4}
 \{q^2\sum_{i=1,2,3,4}
 c_i(\psi^{+f_1}(p)O_i\psi^{f_2}(p_1-q))\times
  $$
  \be
 (\psi^{+f_2}(p_1)O_i\psi^{f_1}(p+q))
 -\sum^5_{i=1} c'_i
 (\psi^{+f_1}(p)\hat q O_i\psi^{f_2}(p_1-q))
 (\psi^{+f_2}(p_1)\hat q O_i\psi^{f_1}(p+q))\}
  \ee
   where $O_i$ and $c_i,c'_i$ are $$
 O_i=1,\gamma_{\mu},\gamma_5\gamma_{\mu},\gamma_5,\sigma_{\mu\nu}=
 \frac{\gamma_{\mu}\gamma_{\nu}-\gamma_{\nu}\gamma_{\mu}}{2i} ~~
 {\rm for}~~ i=1,2,3,4,5
 $$
 \be
 c_i=1,-\frac{1}{2},-\frac{1}{2},-1,0
 \ee
 $$
 c'_i=\frac{1}{4},\frac{1}{4},-\frac{1}{4},\frac{1}{4},\frac{1}{8}
 $$
 Now we take care of the flavour indices
 \be
 \sum^{N^2_f-1}_{F=1}t^F_{fg}t^F_{ed}=-\frac{1}{2N_f}\delta_{fg}
 \delta_{ed}+\frac{1}{2}
 \delta_{fd}\delta_{ge},
 \ee
 to rewrite the curly brackets in (A.9) as
 $$
 \{...\} =2q^2\sum_{i,F}c_i(
 \psi^{+}(p) O_it^F\psi(p_1-q))
 (\psi^{+}(p_1) O_it^F\psi(p+q))+
 $$
 $$
 +\frac{q^2}{N_f}\sum_ic_i(
 \psi^{+}(p) O_i1^F\psi(p_1-q))
 (\psi^{+}(p_1) O_i1^F\psi(p+q))+
 $$
 \be
 -2\sum_{F,i=1}^5c'_i(
 \psi^{+}(p)\hat q O_it^F\psi(p_1-q))
 (\psi^{+}(p_1)\hat q O_it^F\psi(p+q))-
 \ee
 $$
 -\frac{1}{N_f}\sum_{i=1}^5c'_i(
 \psi^{+}(p)\hat q O_i1^F\psi(p_1-q))
 (\psi^{+}(p_1)\hat q O_i1^F\psi(p+q))
  $$

 We now can introduce currents
 $$
  J_i^{F,O}(p)=
  \psi^+(p)
   O_i
(  t^F,  1^F ) \psi(p_1-q)
  $$
 and similarly for $\tilde J_i^{F,O}$, where instead of $O_i$
 stands $\hat q O_i$,
  to rewrite
  all terms in
 (A.12) as
  $$
  \{...\}
 =2q^2\sum_{i,F}c_iJ^F_i(p)J^F_i(p_1)+\frac{q^2}{N_f}\sum_ic_i
 J^0_i(p)J^0_i(p_1)
 $$
 \be
 -2\sum_{i,F}c'_i\tilde
 J^F_i(p)\tilde J^F_i(p_1)-\frac{1}{N_f}\sum_ic'_i
 \tilde J^0_i(p)\tilde J^0_i(p_1)
 \ee

  As a result (A.9) assumes the form
  $$
  <\theta^2_i>=\int \frac{dpdp_1dq}{(2\pi)^{12}}
  \frac{2(4\pi^2\rho^2)^2\varphi^2(q)}{(N_c^2-1)V_4}
 \{2q^2(1-\frac{1}{N_c^2})\sum_{i,F}
 c_iJ^F_i(p)J^F_i(p_1)+
 $$
  $$
  \frac{q^2}{N_f}
 (1-\frac{1}{N^2_c})\sum_{i}c_iJ^{0}_i(p)J^{0}_i(p_1)-2\sum_{i,F}c'_i
 \tilde J^{F}_i(p)\tilde J^{F}_i(p_1)-
 $$
  \be
  -\frac{1}{N_f}
 \sum_{i}c'_i\tilde J^{0}_i(p)\tilde J^{0}_i(p_1) \}.
 \ee

 \end{document}